\documentclass{article}

     \PassOptionsToPackage{numbers, compress}{natbib}

     \usepackage[final]{neurips_2024}

\usepackage[utf8]{inputenc} 
\usepackage[T1]{fontenc}    
\usepackage{hyperref}       
\usepackage{url}            
\usepackage{booktabs}       
\usepackage{amsfonts}       
\usepackage{nicefrac}       
\usepackage{microtype}      
\usepackage{xcolor}         
\usepackage{amsmath}        
\usepackage{graphicx}

\title{\includegraphics[scale=0.03]{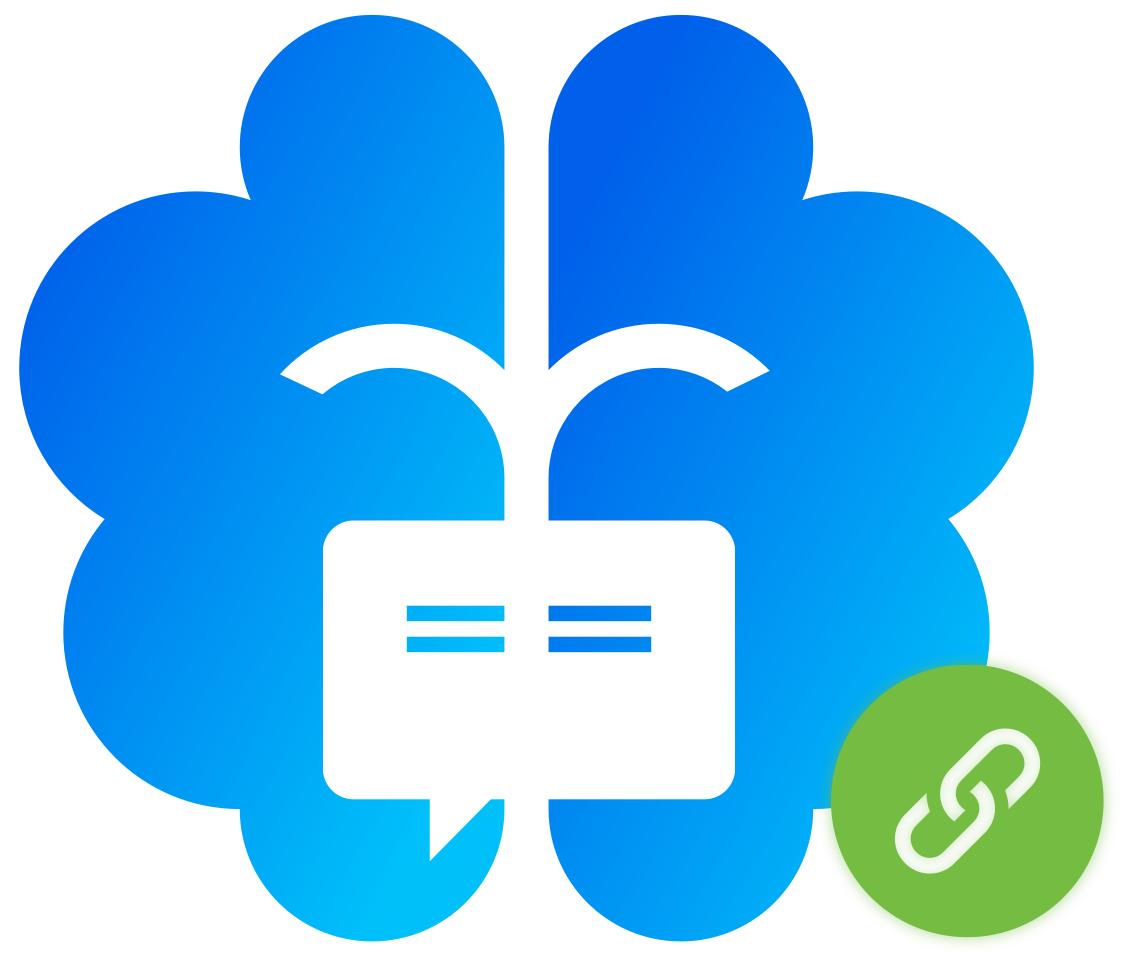}IntelliChain: \\
An Integrated Framework for Enhanced Socratic Method Dialogue with LLMs and Knowledge Graphs
\thanks{Qi, C., Jia L., Wei, Y., Jiang Y.-H., Gu, X. (2024). IntelliChain: An Integrated Framework for Enhanced Socratic Method Dialogue with LLMs and Knowledge Graphs. Conference Proceedings of the 28th Global Chinese Conference on Computers in Education (GCCCE 2024), 116–121. Chongqing, China: Global Chinese Conference on Computers in Education.}}

\author{
Changyong Qi$^{1, 2, 3}$, Linzhao Jia$^{1, 2, 3}$, Yuang Wei$^{1, 2, 3}$, \textbf{Yuan-Hao Jiang}$^{1, 2, 3}$ \textbf{Xiaoqing Gu}$^{4 }$ \thanks{Corresponding Author: xqgu@ses.ecnu.edu.cn} \,\,\\
\\
$^{1}$~\text{Lab of Artificial Intelligence for Education, East China Normal University}\\
$^{2}$~\text{Shanghai Institute of Artificial Intelligence for Education, East China Normal University}\\
$^{3}$~\text{School of Computer Science and Technology, East China Normal University}\\
$^{4}$~\text{Department of Educational Information Technology, East China Normal University}\\
}
\begin{document}
\maketitle
\begin{abstract}
  With the continuous advancement of educational technology, the demand for Large Language Models (LLMs) as intelligent educational agents in providing personalized learning experiences is rapidly increasing. This study aims to explore how to optimize the design and collaboration of a multi-agent system tailored for Socratic teaching through the integration of LLMs and knowledge graphs in a chain-of-thought dialogue approach, thereby enhancing the accuracy and reliability of educational applications. By incorporating knowledge graphs, this research has bolstered the capability of LLMs to handle specific educational content, ensuring the accuracy and relevance of the information provided. Concurrently, we have focused on developing an effective multi-agent collaboration mechanism to facilitate efficient information exchange and chain dialogues among intelligent agents, significantly improving the quality of educational interaction and learning outcomes. In empirical research within the domain of mathematics education, this framework has demonstrated notable advantages in enhancing the accuracy and credibility of educational interactions. This study not only showcases the potential application of LLMs and knowledge graphs in mathematics teaching but also provides valuable insights and methodologies for the development of future AI-driven educational solutions.
\end{abstract}

\section{Introduction}

The advancement of artificial intelligence technology, particularly the development of Large Language Models (LLMs), is propelling educational innovation forward. These models demonstrate significant potential in complex interactions, natural language understanding, and generation, paving the way for new possibilities in creating intelligent educational agents \cite{gan2023large}. In the realm of education, LLMs support more personalized and interactive learning experiences, which are crucial for meeting diverse learning needs and enhancing the quality of education \cite{wang2024large}. However, despite LLMs' remarkable capabilities in enhancing the intelligence of educational applications, they still face challenges regarding accuracy, reliability, and relevance to specific educational content \cite{kaddour2023challenges}.

This study is dedicated to optimizing the information exchange and collaboration mechanisms among intelligent agents through the chain-of-thought dialogue approach, combining LLMs and knowledge graphs, to enhance teaching effectiveness and the learning experience. The chain-of-thought dialogue, an interaction mode that simulates the human thought process, can promote a deeper understanding and response among intelligent agents, offering learners richer and more targeted learning content \cite{chae-etal-2023-dialogue}. However, achieving this goal requires addressing the inherent shortcomings of relying on LLMs, including improving the accuracy of interactions and ensuring the credibility of educational content \cite{bai2023qwen}.

To overcome these challenges, this research proposes an innovative solution: integrating knowledge graphs to enhance the credibility of LLMs' application in educational settings, while optimizing the design and collaboration of the multi-agent system. As an effective tool for organizing and representing information, knowledge graphs provide rich, structured background knowledge for LLMs, aiding intelligent agents in more accurately understanding and responding to educational content \cite{10.1145/3643479.3662055}. Furthermore, by optimizing the design of the multi-agent system, this study aims to achieve more effective collaboration and information sharing among intelligent agents, thereby enhancing the educational outcomes based on the chain-of-thought dialogue approach. The goal of this research is to develop a reliable and effective educational multi-agent system that can leverage LLMs and knowledge graphs to provide personalized learning experiences and promote efficient collaboration among intelligent agents through the chain-of-thought dialogue method. In this way, we aim to offer a new and effective AI application mode in the field of educational technology, especially in supporting complex teaching tasks and facilitating students' learning processes.

\section{Related Work}

Amidst the rapid advancements in artificial intelligence, LLMs have unveiled unprecedented potential in the field of education. Generative models have been demonstrated to effectively support adaptive learning systems, automated content generation, and intelligent assessment mechanisms \cite{bahroun2023transforming}. Their capacity to understand and generate complex natural language offers robust support for personalized educational pathways. However, the academic community has extensively discussed the limitations and accuracy challenges of LLMs when dealing with domain-specific knowledge \cite{yao2023knowledge}. Despite their revolutionary capabilities in language processing, how to effectively leverage these capabilities to enhance learning efficiency and outcomes remains an open question in educational applications.

\subsection{Chain-of-Thought Dialogues in Educational Technology}

Chain-of-thought dialogue, as a method simulating human thought processes, has gained considerable attention in educational technology research in recent years. By facilitating deeper and more coherent exchanges, this approach aims to promote critical thinking and deep understanding among learners. Research has highlighted that the implementation of chain-of-thought dialogue significantly enhances learning motivation and engagement in online learning environments \cite{wang2023cue}. The integration of LLMs in chain-of-thought dialogue represents a significant leap forward in educational technology research. These models, when applied to simulate and generate chain-of-thought dialogues, offer the potential to facilitate more nuanced and deep learning interactions \cite{yin2023exchange}. Recent studies have explored how LLMs like GPT-3.5 can be fine-tuned to produce sequential reasoning steps in problem-solving tasks, effectively mimicking the human thought process \cite{liu2023improving}. This application not only enhances the dialogue's relevance and depth but also promotes a more engaging and interactive learning experience. However, despite these advancements, challenges persist in ensuring the dialogues' consistency with factual accuracy and pedagogical objectives, indicating a fertile area for future research.

\subsection{Leveraging Knowledge Graphs and Multi-Agent Systems to Transform Educational Interactions}

Knowledge graphs, as a structured representation of knowledge, play a critical role in enhancing the knowledge base and reasoning capabilities of LLMs. By integrating rich domain knowledge in the form of graphs, LLMs can achieve more accurate contextual understanding and reasoning within educational settings, significantly improving the quality and accuracy of responses \cite{gan2023large}. Additionally, knowledge graphs provide detailed entity relationships and attribute information, effectively aiding LLMs in avoiding the generation of hallucinations—content that appears plausible but is factually incorrect \cite{agrawal2023can}. This capability is especially important for educational applications, as it directly relates to the reliability of learning content and the quality of education. Simultaneously, the development of multi-agent systems based on LLMs represents a new zenith in intelligent teaching and interactive learning within the field of educational technology. Agents within these systems leverage the computational power and advanced language processing abilities of LLMs to facilitate efficient communication and collaboration amongst themselves. More importantly, they can continuously learn from interactions with learners, thereby dynamically adapting to learners' needs and providing personalized learning support. Furthermore, by simulating complex interactions of the real world, multi-agent systems offer learners a more rich and immersive learning experience. The flexibility and dynamism of these systems are unparalleled by traditional educational methods.

Integrating knowledge graphs and multi-agent systems based on LLMs, especially in supporting chain-of-thought dialogues based on LLMs, not only enhances the accuracy and relevance of educational content but also greatly increases the personalization and interactivity of educational interactions. This transforms the educational process from a one-way transmission of knowledge into a dynamic, bidirectional, and interactive learning experience.

\section{Framework}

\begin{figure}[t]
  \centering
  \includegraphics[width=\linewidth]{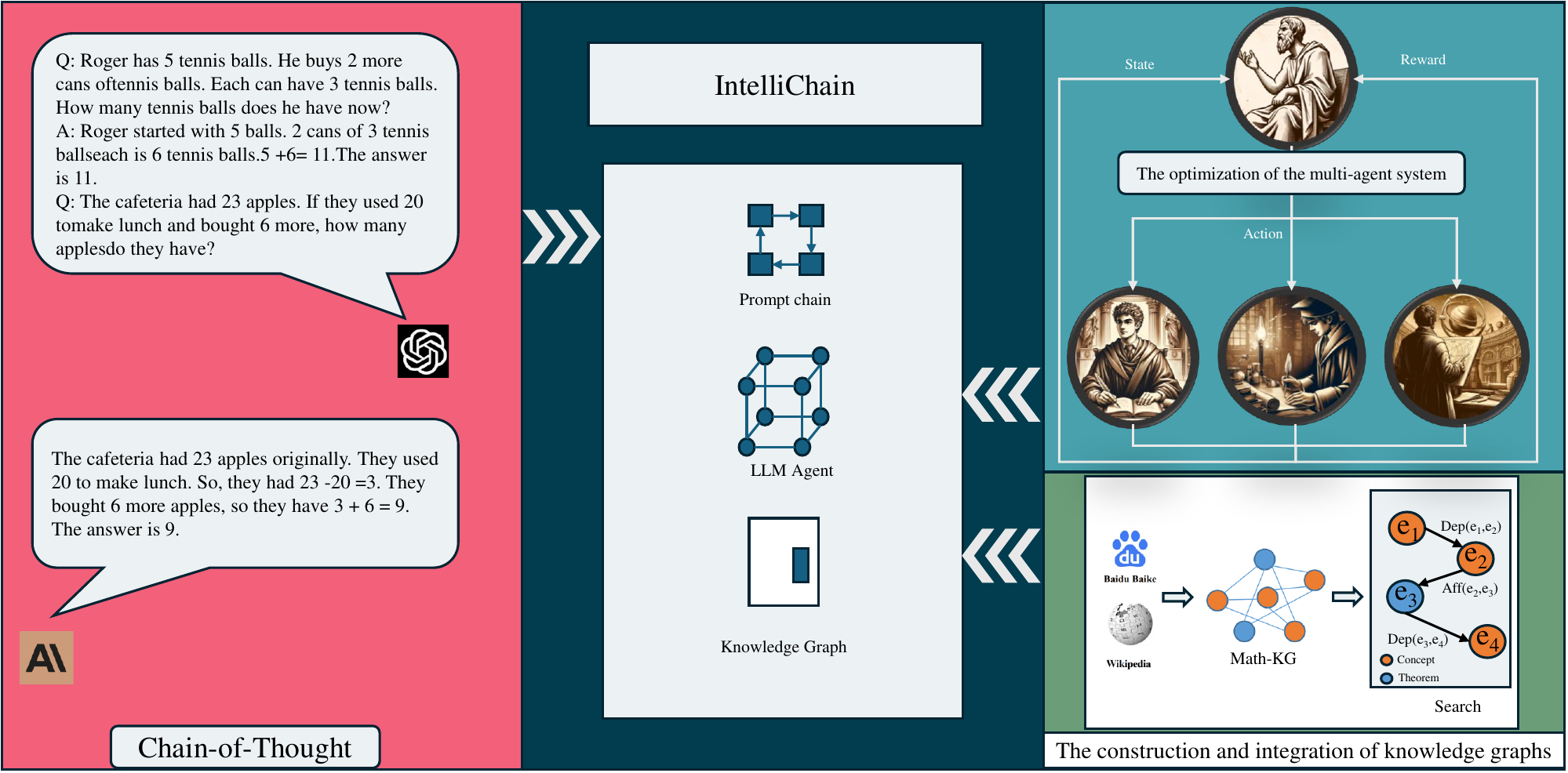}
  \caption{The Framework of IntelliChain.}
  \label{fig:1}
\end{figure}

In this study, we introduce IntelliChain, a comprehensive educational support architecture that synergistically integrates LLMs, knowledge graphs, and a multi-agent system to facilitate an efficient chain-of-thought dialogue mechanism, as shown in Figure \ref{fig:1}. Specifically, IntelliChain is built upon modular principles with a focus on optimizing the collaborative efficiency of intelligent agents in three main areas: the strategy design of chain-of-thought dialogue, the construction and integration of knowledge graphs, and the optimization of the multi-agent system. This framework not only aims to enhance the accuracy and relevance of educational content but also leverages the unique strengths of each component to support nuanced educational dialogues, thereby setting a new standard for interactive learning environments.

Within the IntelliChain framework, an innovative dialogue strategy is articulated, leveraging a sophisticated chain-of-thought approach to enhance the pedagogical efficacy of educational dialogues. This methodology is underpinned by an integration of advanced pedagogical principles, encompassing strategies such as guided questioning, sequential analytical reasoning, iterative feedback mechanisms, and the facilitation of exploratory inquiry. The design premise of IntelliChain advocates for a structured dialogue progression, where learners are methodically navigated through complex cognitive tasks, thereby augmenting their analytical acumen and fostering critical thinking capacities. The framework introduces a novel role-based interaction schema, wherein agents designated as "instructors" and "learners" partake in a reciprocal exchange of investigative queries and insights. This orchestrated educational dialogue not only promotes active learner engagement through sustained cognitive involvement but also ensures pedagogical alignment via continuous feedback loops.

The construction and integration of the knowledge graph represent a cornerstone in facilitating enriched educational dialogues, specifically tailored to the domain of mathematics education. This process is undergirded by a meticulous amalgamation of domain expertise and high-caliber educational resources, culminating in the development of a knowledge graph that meticulously catalogues key mathematical concepts, principles, and illustrative examples. The resultant knowledge base serves as a robust and comprehensive repository, furnishing LLMs with precise and extensive domain-specific information. To achieve an efficacious integration of the knowledge graph with LLMs, this study introduces an advanced querying mechanism designed to harness the knowledge graph's potential fully. Prior to each dialogue iteration, the system autonomously conducts a query within the knowledge graph based on the specific knowledge points implicated in the dialogue content. This procedure extracts pertinent information to serve as prompts for dialogue generation, thereby ensuring the educational content's relevance and accuracy.

The optimization of the multi-agent system represents a strategic deployment of intelligent agents, precisely tailored to the specific demands of educational tasks. This strategy is designed to fully leverage the unique expertise of each agent, thereby enhancing the quality of instructional content and the efficiency of the learning process. At the heart of system optimization lies the adoption of advanced reinforcement learning algorithms, establishing a learning and adaptation mechanism. This mechanism enables intelligent agents to iteratively adjust their behaviors and strategies based on feedback received from interactions with learners. This adaptive process allows the system to self-modulate based on the correlation between the agents' actions and learners' responses, ensuring that educational interventions are optimally aligned with the learners' evolving needs and preferences.

\section{Results}

\begin{figure}[t]
  \centering
  \includegraphics[width=\linewidth]{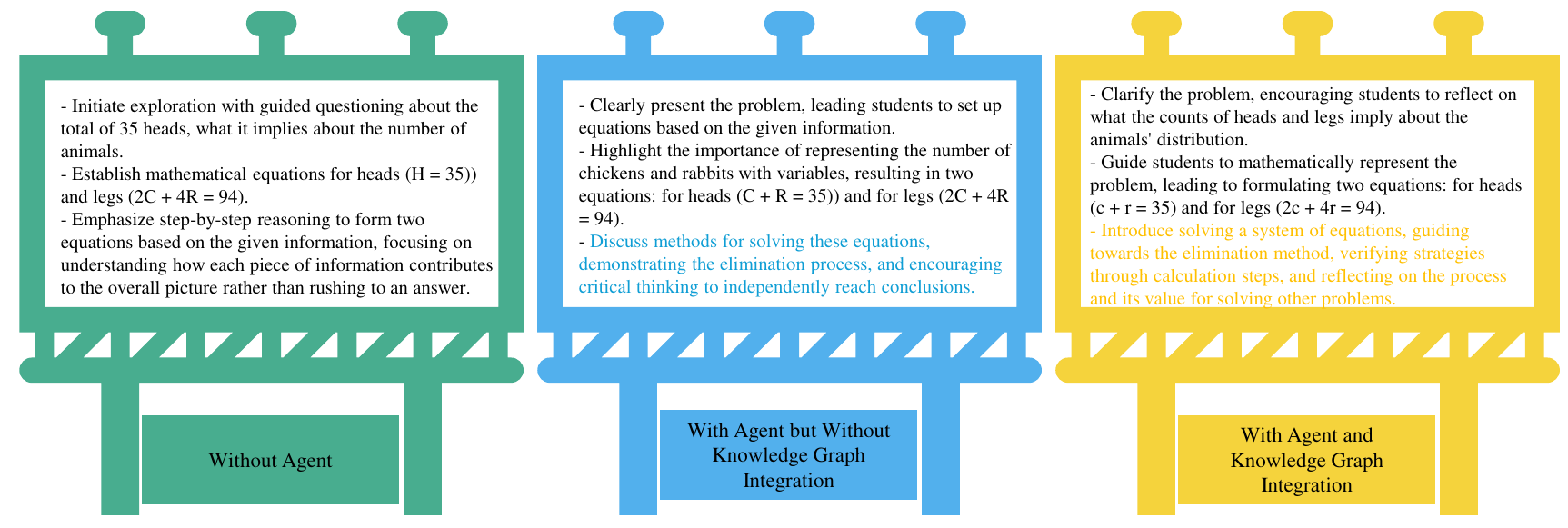}
  \caption{ Differential Teaching Outcomes Using the IntelliChain Framework Across Three System Configurations.}
  \label{fig:2}
\end{figure}

In the present study, the IntelliChain framework was utilized to examine the differential outputs of teacher agents employing the Socratic method in solving the classical chicken-rabbit problem under three distinct system configurations: without agents, with agents but without knowledge graph integration, and with agents integrated with a knowledge graph, as shown in Figure \ref{fig:2}. This comparative analysis aimed to elucidate the impact of knowledge graph integration and multi-agent system optimization on the quality of pedagogical dialogue and teaching efficacy.

Without Agent Configuration: In the absence of agent intervention, the teaching dialogue primarily relied on open-ended questioning and step-by-step reasoning to stimulate learner exploration. While this approach facilitated learner engagement, it lacked the specificity and efficiency that could be achieved through the utilization of specific algorithms or educational resources, rendering the teaching process somewhat generic and less targeted. With Agents but Without Knowledge Graph: The configuration of employing teaching agents without knowledge graph integration showed a marked improvement in teaching dialogue by adhering to the Socratic method, characterized by structured problem presentation and algebraic problem-solving guidance. Despite the potential of agents in enhancing the learning process, the absence of knowledge graph support limited the depth and breadth of instructional content, underutilizing the potential benefits of agent involvement in teaching. With Agents Integrated with Knowledge Graph: The integration of teaching agents with a knowledge graph significantly enhanced the quality of teaching dialogue. This configuration not only enabled agents to guide the problem-solving process through algebraic methods effectively but also deepened the discussion on problem context and related concepts utilizing the rich information from the knowledge graph. Such an integrated approach not only deepened learners' understanding of the problem but also promoted personalized learning and active learner participation by dynamically adjusting teaching strategies and methods.

\section{Discussion}

The IntelliChain framework constitutes a pivotal technological advancement in the realm of educational technology, particularly in augmenting Socratic method through the integration of chain-of-thought dialogues, knowledge graphs, and an optimized multi-agent system. This comprehensive strategy not only facilitates a deeper and structured inquiry, akin to the Socratic method, but also pioneers a personalized educational pathway via adaptive learning strategies. Despite the framework's promising capabilities, challenges such as its applicability across diverse learning scenarios and the imperative for maintaining up-to-date, unbiased knowledge graphs warrant further investigation. Future research directions might include refining the framework's dialogue strategies to accommodate various learning styles and broadening the scope of knowledge graphs to cover a wider spectrum of disciplines. Additionally, the potential integration of emergent technologies presents an exciting frontier for creating more immersive and interactive learning environments, underscoring IntelliChain's transformative potential in educational methodologies.

\section{Conclusion}

This study introduced the IntelliChain framework, an innovative approach that enhances educational dialogues through the integration of LLMs, knowledge graphs, and a multi-agent system, optimized for the Socratic method. It demonstrated the framework's capability to improve the precision and relevance of educational content while facilitating personalized learning experiences. Comparative analysis across different configurations underscored the significance of knowledge graph integration and multi-agent system optimization in augmenting teaching efficacy. Despite promising outcomes, challenges such as adaptability across diverse learning scenarios and maintaining unbiased knowledge graphs remain. Future efforts will focus on refining dialogue strategies and exploring emerging technologies to further enhance the learning environment. IntelliChain represents a significant stride towards advancing AI-driven personalized education, promising to elevate the quality and effectiveness of teaching and learning methodologies.

\bibliographystyle{IEEEtran}
\bibliography{sample-base}

\end{document}